\documentclass[twocolumn,showpacs,preprintnumbers,amsmath,amssymb]{revtex4}
\usepackage{graphicx}

\usepackage{amssymb}
\begin{document}
%\preprint{APS/123-QED}
\draft

\title{Semiflexible Polymer Confined in Closed Spaces}
\author{Takahiro Sakaue}
 %\altaffiliation[Also at ]{Yukawa Institute for Theoretical Physics, Kyoto University.}%Lines break automatically or can be forced with \\
\email{sakaue@scphys.kyoto-u.ac.jp}
\affiliation{
Department of Physics, Graduate School of Science, Kyoto University, Kyoto 606-8502, Japan\\
}%

%\affiliation{

\date{\today}% It is always \today, today,

\date{\today}% It is always \today, today,

\begin{abstract}
We analyze static properties of a strongly confined semiflexible polymer, i.e. either trapped in a closed space or compressed by external forces, in an athermal solvent.
Like a flexible polymer case, we can resort to an analogy with the semidilute solution, but a complication due to the additional length scale arising from the chain rigidity results in more diverse behaviours depending on system parameters.
For each regime, scaling forms of the excess free energy of the confinement are derived.
Effects of the confinement geometry and the solvent quality are briefly discussed.
%Among others, ideal chain behaviours of a confined chain are expected (excluded volume effect is disregarded) at certain conditions as a consequence of the chain rigidity and the confinement.
\end{abstract}


\maketitle

\section{Introduction}
\label{Introduction}
Macromolecules are very soft objects. 
As a consequence, they can be squeezed into spaces with spatial size substantially smaller than their preferable size in bulk~\cite{Cassasa,Daoud_deGennes,deGennes,deGennes_review}.
Understanding behaviours of such confined polymers serves as the basis for range of phenomena encountered in polymer industry, biotechnology and various molecular processes in living cells.
Thanks to the progress in nanofabrication techniques and single molecule observations, experiments in this filed are now rapidly developing~\cite{Kasianowicz,DrivenDNA,DNA_partitioning_model_pore,Ichikawa,CompressionDNA,DNAinNonochannels}.

%One of the most important quantities on confined polymer is the excess free energy of the confinement, i.e., the work required to achieve the confinement.
We have recently pointed out that polymer chains can be intensively compressed in certain situations so that the excess free energy of the confinement $F$ exhibits a nonlinear dependence on the chain length $N$.
This led us to propose two distinct confinement regimes, i.e., {\it weak confinement} (WCR) and {\it strong confinement} (SCR)~\cite{Sakaue_confinement}.
The qualitative difference between these two are easily understood using the example of a flexible linear chain.
Suppose that a long flexible chain with the bulk size $R \simeq b N^{3/5}$ ($b$ is the monomer size) is confined between the parallel plates with their separation $D<R$.
In a length scale smaller than $D$, the behaviours of the chain is not perturbed by the wall, and this introduces the blob size $D$, inside which there are $g=(D/b)^{3/5}$ monomers.
The local monomer volume fraction inside blob $\phi \simeq b^3 g/D^3 \simeq (b/D)^{3/4}$ is {\it independent} of the chain length.
This corresponds to what we call WCR.
The extensive nature of the free energy, therefore, results in the conventional scaling form of the confinement free energy $F \simeq (b/D)^{3/5} N$, that is proportional to the chain length~\cite{Daoud_deGennes,deGennes}.
The same argument is applicable for the chain confined in the capillary, too.
However, it is no longer hold for the chain trapped in the closed cavity, where the volume fraction increases with the chain length~\cite{RedBook}.
This provides the simplest example of the SCR, in which several distinct properties are expected~\cite{Sakaue_confinement}.

In ref.~\cite{Sakaue_confinement}, we generalized this concept to the branched polymers, where the constraint imposed by the connectivity results in nontrivial confinement behaviours.
Another requisite extension is examining the effect of the {\it stiffness}.
The reason is immediately recognized by the fact that many biopolymers such as DNA and various proteins have a quite large stiffness aspect ratio $p=l/b$, where $l$ is the Kuhn segment length.
Even for less stiff chains than DNA (note that any real polymers possessed some degree of intrinsic stiffness), such a consideration may become apparent in denser system, in particular in the confined system, where the additional length scale of the confinement $D$ would produce a variety of the behaviours.

It would be convenient to introduce the distinction between WCR and SCR for semiflexible chain, too.
Scaling laws describing properties of a semiflexible linear chain trapped in the capillary or the slit (WCR) are the same as those of a flexible one if the size of the confinement is much larger than the persistence length $l_p (\sim l)$; the effect of the stiffness shows up just through the definition of link size.
At $D<l$, however, the chain is obliged to be bent sharply.
This leads to the introduction of the Odijk deflection length $l_{\rm O} \simeq D^{2/3}l_p^{1/3}$, which provides $l$ dependences of various quantities~\cite{Odijk}.
%  qualitative understanding is possible thanks to the introduction of the Odijk deflection length $l_O \simeq (Dl_p)^{1/3}$.
Recent observation of a single DNA chain in a narrow channel have shown the crossover from a flexible to a semiflexible chain regime around the channel size $D \simeq l$~\cite{DNAinNonochannels}.

A congeneric level of argument seems to be lacking for the semiflexible linear chain in SCR, e.g. in a closed cavity, and it is the purpose of the present paper to provide a qualitative guide for such a problem.
%The purpose of the present paper is to provide a qualitative guide on the semiflexible linear chain in SCR, i.e. in a closed cavity.
In particular, we discuss scaling forms of the confinement free energy, which is one of the most important quantities in the problem, with their physical origin in various regimes.
To do so, we first discuss the static property of a flexible linear polymer in SCR in some details in Sec. \ref{flexible}.
Based on the semidilute solution analogy, we show that the confinement free energy scales $F \sim N^{9/4}$ under good solvent condition~\footnote{In good solvent regimes, the exact critical exponent is known to be $\nu= 0.588 \pm 0.001$ due to the correlation effect~\cite{DoiEdwards} and this results in $F \sim N^{2.309}$. In the present paper, we use the mean field exponent $\nu=3/5$ for simplicity.}, while $F \sim N^3$ at ``lower" temperature and high concentration .
Then, in Sec. \ref{semiflexible}, we proceed to the more general case of a semiflexible linear polymer.
As we shall show, the presence of multiple length scales, i.e., the cavity size, the Kuhn length and the correlation length of the concentration fluctuation, leads to rich confinement regimes even in the athermal solvent.
Some of these can be viewed as a semidilute solution as in the flexible chain case, and in such cases, we can directly apply the result known in the theory of semidilute solution of semiflexible chains~\cite{Schaefer,RedBook}.
However, there are other regimes, which manifest intrinsic effects of the confinement originated from either a reduction in the conformational entropy or a sharp bending.
Throughout the article, we assume that the segment concentration of the confined chain is not very high, in which the model independent universal behaviours can be discussed.
To make the discussion simple, we also assume that the confinement geometry is a sphere (the effect of the shape will be addressed in Sec. \ref{Discussion}).

\section{A flexible chain in SCR}
\label{flexible}
Let us consider a flexible linear chain (contour length $L$) dissolved in a solvent.
Large-scale properties of the chain is conveniently modelled as a sequence of $N$ links of size $b$ (volume $v \simeq b^3$), i.e., the standard bead-on-filament model, with the distance between neighboring beads along the chain $a$~\cite{RedBook}.
For a flexible chain, the natural choice is $N=L/b$ and $a=b$.
The excluded volume effect becomes important when the parameter $z(N) = B N^{1/2}/a^3$ is larger than unity, where $B = v \tau$ is the second virial coefficient.
The solvent quality is designated by a reduced temperature $\tau = (T-\theta)/\theta$ with $T$ and $\theta$ being the absolute temperature and a $\theta$ temperature, respectively.
From the condition $z(g_{th})=1$, the number of monomers $g_{th}=\tau^{-2}$ constituting the thermal blob is obtained.
Inside the thermal blob, the chain conformation is Gaussian, leading to the thermal blob size $\xi_{th} = b \tau^{-1}$.
%If the solvent is not the athermal, the repulsion due to the excluded volume effect is compensated to some degree.
The chain conformation in a bulk solution can be envisioned as a self-avoiding walk of thermal blobs of size $\xi_{th} \simeq b/ \tau$, which leads to the overall chain size $R \simeq b N^{3/5}  \tau^{1/5}$.

Then, consider that the chain is brought into a spherical cavity of size $D < R$. 
This corresponds to the simplest example of the SCR.
A key observation is that the strongly confined chain can be viewed as a semidilute solution with the volume fraction $\phi \simeq b^3N/D^3$~\cite{Sakaue_confinement}, i.e., the correlation of concentration fluctuations is suppressed at the length scale 
\begin{eqnarray}
\frac{\xi}{b} \simeq  \phi^{-\frac{3}{4}} \tau^{-\frac{1}{4}} \simeq \left(\frac{D}{b} \right)^{\frac{9}{4}}N^{-\frac{3}{4}} \tau^{-\frac{1}{4}}
\label{xi_good}
\end{eqnarray}

The above expression for the correlation length is obtained from a scaling argument~\cite{deGennes} (by imposing $\xi$ must be (i) independent of $N$ and (ii) equal to $R$ at the overlap concentration) and valid as long as $\phi<\tau \iff \xi > \xi_{th}$.
This corresponds to the {\it fluctuating semidilute regime}.
Inside the blob, the correlation is evident, thus, the number of monomers in each blob is obtained from the relation $\xi \simeq b g^{3/5} \tau^{1/5}$.
At larger length scale, the application of the mean-field argument is valid with the blobs of size $\xi$ being the renormalized monomers.
A free energy arising from volume interactions is , thus, evaluated as 
\begin{eqnarray}
\frac{F_{vol}}{k_BT} \simeq B^* \left\{\frac{(N/g)^2}{D^3}\right\} \simeq N \phi^{\frac{5}{4}} \tau^{\frac{3}{4}} \simeq \left( \frac{b}{D}\right)^{\frac{15}{4}}N^{\frac{9}{4}} \tau^{\frac{3}{4}}
\label{F_good}
\end{eqnarray}
 where $B^* \simeq \xi^3$ is the renormalized second virial coefficient. 
This is equivalent to assigning the energy on the order of $k_BT$ to each blob.

\begin{figure}
\includegraphics[width=7.5cm]{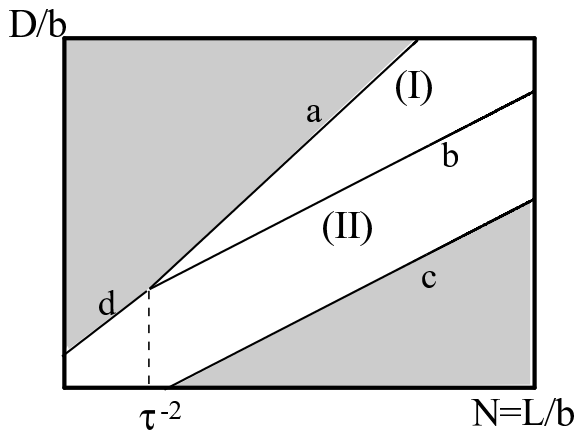}
\caption{A diagram of confinement regimes for a flexible chain in a closed sphere (logarithmic scale). Region (I) and (II) correspond to {\it fluctuating semidilute} and {\it semidilute in $\theta$-solvent} regime, respectively. Shaded areas are irrelevant, i.e., upper left is a bulk ($D>R$) and lower right is not accessible ($\phi \ge 1$).  Border lines between regimes are (a) $D/b \simeq \tau^{1/5}N^{3/5}$, (b) $D/b \simeq (N/\tau)^{1/3}$, (c) $D/b \simeq N^{1/3}$, (d) $D/b \simeq N^{1/2}$.}
\label{diagram_flexible}
\end{figure}

In the opposite case, $\phi>\tau  \iff \xi < \xi_{th}$, the chain conformation is Gaussian in all the length scale, which indicates the applicability of the mean field theory.
This corresponds to the {\it semidilute in $\theta$-solvent regime}.
The volume interaction energy per unit volume can be expressed in a virial expansion $f_{vol}/k_BT \simeq 1/2 Bc^2 +1/3 Cc^2 + $, where $B= v \tau$, $C= v^2$ are the second and third virial coefficients and $c=\phi/v$ is the monomer concentration.
The monomer pair-correlation function is calculated by a random phase approximation~\cite{DoiEdwards,deGennes,RedBook}
\begin{eqnarray}
\langle \delta c (\mathbf{r}) \delta c (0) \rangle = \frac{3c}{2 \pi a^2 r}\exp{\left(-\frac{r}{\xi_{MF}}\right)}
\end{eqnarray}
where, the correlation length in this regime is, instead of eq. (\ref{xi_good}), given by
%\begin{eqnarray}
%\xi^2 = \frac{a^2}{12(Bc+Cc^2)} 
%\label{xi_theta_formal}
%\end{eqnarray}
\begin{eqnarray}
\xi_{MF}^2 = \frac{a^2}{12} \left( \frac{1}{k_BT}\frac{\partial \Pi}{\partial c}\right)^{-1}
\label{xi_theta_formal}
\end{eqnarray}
where the osmotic pressure (due to volume interactions) can be expanded in a series of power $c$ as $\Pi /(k_BT) = B c^2 + 2 C c^3 + \cdots$.
Under the present condition $\phi>\tau$, the interaction is dominated by the three body contribution, therefore, the total volume interaction contribution to the confinement free energy and the correlation length are evaluated as 
\begin{eqnarray}
\frac{F_{vol}}{k_BT} \simeq Cc^3 D^3 \simeq \left(\frac{b}{D}\right)^6 N^3
\label{F_theta}
\end{eqnarray}
\begin{eqnarray}
\frac{\xi_{MF}}{b} \simeq \phi^{-1} \simeq \left( \frac{D}{b}\right)^3 N^{-1} 
\label{xi_theta}
\end{eqnarray}
Note that short chains ($N <g_{th}$) are in the $\theta$ solvent regime already in the bulk isolated state, to which  eq. (\ref{F_theta}) and (\ref{xi_theta}) always hold upon the confinement.
The confinement regimes of a flexible chain in a closed cavity is presented in diagrammatic form in Fig. \ref{diagram_flexible}.

The scaling form of eq.~(\ref{F_good}) in the athermal solvent has  been confirmed by a recent computer simulation~\cite{Luijten}, while the conjecture has been proposed~\cite{Sakaue_confinement} that the form of eq.~(\ref{F_theta}) would be relevant to resolve controversial experimental results on the partitioning coefficient of a flexible polymer into a protein pore~\cite{Kasianowicz,aHL,aHL2,aHL3}.

In both regimes, aside from the volume interaction free energy (bulk term), there is a contribution from the nonuniform link concentration $\Delta F$ (surface term)~\cite{Sakaue_confinement}.
These are calculated in Appendix~\ref{DeltaF} and given by
\begin{eqnarray}
\frac{\Delta F}{k_BT} \simeq \phi^{\frac{3}{2}} \tau^{\frac{1}{2}}\left(\frac{D}{b}\right)^2 \simeq \left(\frac{b}{D} \right)^{\frac{5}{2}} \tau^{\frac{1}{2}}N^{\frac{3}{2}}
\label{DeltaF_good}
\end{eqnarray}
in the fluctuating semidilute regime and
\begin{eqnarray}
\frac{\Delta F}{k_BT} \simeq \left(\frac{\phi D}{b}\right)^2 \simeq \left(\frac{b}{D} \right)^{4}N^{2}
\label{DeltaF_theta}
\end{eqnarray}
in the semidilute in $\theta$-solvent regime.

%One arises from the concentration gradient at the surface region, which results in the  loss of the conformational entropy (Lifshitz entropy) $F_{surf}/k_B T\simeq D^2/\xi^2$.
%The other is associated with the fact that a long chain is reflected at the cavity wall after every $g_D$ monomers.
%This is, in fact, the same origin as the confinement free energy of an ideal chain, and the number $g_D$ is obtained from $\xi (g_D/g)^{1/2} =D$ and $b g_D^{1/2}=D$ for respective regimes.
%The associated reduction in the entropy is $F_{ent}/k_BT \simeq N/g_D$. 
For a flexible chain, the dominant term is always given by the volume interaction, i.e., $F \simeq F_{vol}$.
However, as we shall see below, there is a regime, in which $\Delta F$ plays a central role for a semiflexible chain.

\section{A semiflexible chain in SCR}
\label{semiflexible}
\begin{figure}
\includegraphics[width=7.5cm]{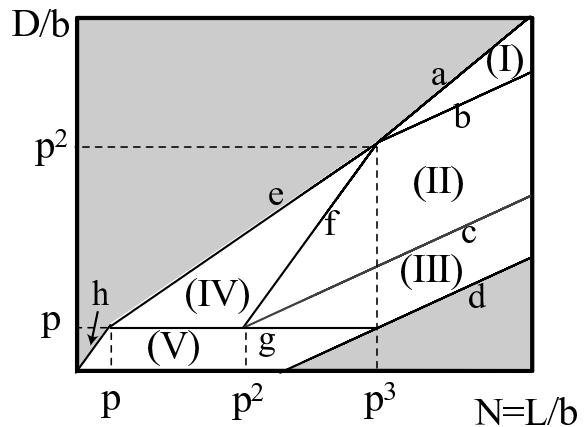}
\caption{A diagram of confinement regimes for a semiflexible chain in a closed sphere under athrmal condition (logarithmic scale). Each labelled region designates (I) {\it fluctuating semidilute}, (II) {\it mean-field semidilute}, (III) {\it liquid crystalline}, (IV) {\it ideal chain} and (V) {\it bending} regime, respectively (see the main text for details). Shaded areas are irrelevant, i.e., upper left is a bulk ($D>R$) and lower right is not accessible ($\phi \ge 1$). Border lines between regimes are (a) $D/b \simeq p^{1/5}N^{3/5}$, (b) $D/b \simeq p N^{1/3}$, (c) $D/b \simeq (pN)^{1/3}$, (d) $D/b \simeq N^{1/3}$, (e) $D/b \simeq (pN)^{1/2}$, (f) $D/b \simeq N/p$, (g) $D/b \simeq p$, (h) $D/b \simeq N$.}
\label{diagram_stiff}
\end{figure}

Our result is summarized in the diagram of Fig. \ref{diagram_stiff}.
To discuss this diagram, we first assume (i) $D>l$ and (ii) $L>pD$ (here $p$ is the stiffness ratio mentioned in Sec. \ref{Introduction}).
As we shall see later, these conditions are requisite for a confined chain to be viewed as an analogue of bulk semidilute solution. 
%The first condition means that the size of the confinement is large enough so that the bending energy is not significant.
%The second one is not obvious for the moment and shall be clarified later.

Let us begin with the evaluation of the volume interaction parameter $z=BN^{1/2}/a^3$.
There is a freedom on how to break the chain into links, and we adopt the simplest, in which the chain is represented as $N=L/b$ links with their volume $v\simeq b^3$ ($b$ is the chain thickness). Then, for the model to be consistent, the distance between links should be set as $a=(bl)^{1/2}$~\cite{RedBook}.
It should be emphasized that results obtained are independent of such a modelling (one can also set $N=L/l$, $v \simeq b l^2$ and $a=l$).
For simplicity, we shall focus on the purely repulsive case, i.e., athermal solution: $\tau =1$ (thus, $B=v$).
From $z=1$, we find the length scale $\xi_{th} \simeq bp^2$, below which the excluded volume effect is insignificant (the corresponding number of monomers is $a g_{th}^{1/2} \simeq \xi_{th}$): for semiflexible chains, $\xi_{th} \gg b$ even in the athermal limit.
As a consequence, the chain with $N<p^3$ behaves as a Gaussian chain, i.e., its size in bulk is given by $R \simeq (lL)^{1/2}$.
On the other hand, the excluded volume interaction should be taken into account for the longer chain, i.e. $R \simeq \xi_{th} (N/g_{th})^{3/5} \simeq (bl)^{1/5}L^{3/5}$.

Along with the same argument as in the flexible chain, one can distinguish (I) {\it fluctuating semidilute regime} and (II) {\it mean-field semidilute regime} depending on the relative ratio of $\xi_{th}$ and the correlation length of density fluctuations.

(I) {\it Fluctuating semidilute regime}: $pN^{1/3}<D/b<(p)^{1/5}N^{3/5}$

According to the analogy with the semidilute solution, a confined chain is viewed as a dense packing of blobs of size $\xi=\xi_{th} (g/g_{th})^{3/5}$ ($g$ is the number of monomers in each blob).
This indicates the condition $g/\xi^3 = N/D^3$, from which one finds
\begin{eqnarray}
\frac{\xi}{b} \simeq \phi^{-\frac{3}{4}} p^{-\frac{1}{4}} \simeq \left(\frac{D}{b} \right)^{\frac{9}{4}} p^{-\frac{1}{4}}N^{-\frac{3}{4}}
\label{xi_fluct_stiff}
\end{eqnarray}
The dominant term in the confinement free energy comes from the volume interaction
\begin{eqnarray}
\frac{F}{k_BT} \simeq \frac{F_{vol}}{k_BT} \simeq \frac{D^3}{\xi^3}\simeq \left(\frac{b}{D}\right)^{\frac{15}{4}}p^{\frac{3}{4}}N^{\frac{9}{4}}
\end{eqnarray}

(II) {\it Mean-field semidilute regime}: $(pN)^{1/3}<D/b<\min\{pN^{1/3},p^{-1}N\}$

A correlation length $\xi$ becomes shorter with the increase in $L$ (or decrease in $D$) and becomes comparable to $\xi_{th}$ at $D\simeq lN^{1/3} (\iff \phi \simeq p^{-3}$).
This indicates a crossover to the mean-field semidilute regime.
One can also recognize the meaning of this condition through the Ginzburg criterion:
\begin{eqnarray}
V_{\xi}^{-2}\int_{V_{\xi}}d^3 {\mathbf r} d^3{\mathbf r' }\langle \delta c({\mathbf r}) \delta c({\mathbf r'}) \rangle  &<&  c^2 \\
\Leftrightarrow
\left( \frac{1}{k_BT}\frac{\partial \Pi}{\partial c}\right)^{\frac{1}{2}} &\lesssim& cv p^{\frac{3}{2}} \label{Gcriterion} \\
\Leftrightarrow
p^{-3} &\lesssim& \phi
\end{eqnarray}
where bracket denotes the statistical average and $V_{\xi} \simeq \xi_{MF}^3$ is the correlation volume.
The correlation length is calculated from eq. (\ref{xi_theta_formal}) with ignoring the third and higher order terms in virial expansion
\begin{eqnarray}
\frac{\xi_{MF}}{b} \simeq \left(\frac{p}{\phi} \right)^{\frac{1}{2}} \simeq \left(\frac{D}{b} \right)^{\frac{3}{2}} \left(\frac{p}{N} \right)^{\frac{1}{2}} 
\label{xi_mean_field}
\end{eqnarray}
(Note that the eq.~(\ref{Gcriterion}) also provides the borderline for a flexible chain discussed in the previous section.)

 %which is calculated to be $\phi \gtrsim \tau p^{-3}$ for semiflexible polymer solutions (note that the limit of $p \rightarrow 1$ gives the borderline for a flexible chain discussed in the previous section).
Here again, the confinement free energy is dominated by the volume interaction, which is evaluated by the mean field theory.
In the case under consideration ($(blL)^{1/3}<D$), the binary interaction is the most dominant than higher order terms in the virial expansion, thus,
\begin{eqnarray}
\frac{F}{k_BT} \simeq \frac{F_{vol}}{k_BT} \simeq B \left(\frac{N^2}{D^3}\right)\simeq \left( \frac{b}{D} \right)^3 N^2
\label{F_meanfield}
\end{eqnarray}
This scaling form has been proposed on the basis of a self-consistent field calculation~\cite{Muth3}, but here one should notice the range for the applicability, i.e. only valid for a semiflexible chain in the present regime.
%From eq. (\ref{xi_theta_formal}), the correlation length is given by
%\begin{eqnarray}
%\frac{\xi}{b} \simeq \left(\frac{p}{\phi} \right)^{\frac{1}{2}} \simeq \left(\frac{D}{b} \right)^{\frac{3}{2}} \left(\frac{p}{N} \right)^{\frac{1}{2}} 
%\label{xi_mean_field}
%\end{eqnarray}

(III) {\it Liquid crystalline regime}: $p<D/b<(pN)^{1/3}$

If the density is further increased, the volume interaction becomes higher and the correlation length becomes comparable to the Kuhn segment length at $D \simeq (blL)^{1/3} (\iff \phi \simeq p^{-1}$).
This indicates some role of model specificity on the flexibility mechanism at higher concentrations, analysis of which is beyond the scope of the present discussion.
Here, we just note what is expected from the bulk theory.
According to the standard theory~\cite{Gresberb_Khokhlov_LC}, the system responds by breaking the isotropic symmetry, i.e., the competition between steric repulsion (evaluated by the second virial approximation) and orientational entropy results in the first-order phase transition at the volume fraction $\phi \simeq p^{-1}$.
The transition proceeds with the coexistence of isotropic and nematic phases at $c_1 p^{-1} < \phi <c_2 p^{-1}$ (numerical coefficients depend on the flexibility mechanism), which suggests an interesting possibility of an ``intra-chain" separation for a confined chain.
%More precisely, the semiflexible polymer solution is isotropic at $\phi <c_1 p^{-1}$ and highly oriented (nematic) at $\phi >c_2 p^{-1}$.
%Between these two concentrations, the solution exhibits a phase separation.
%To reach this conclusion, the translational entropy of individual semiflexible chains plays no role, therefore, the results can be applicable to our confined chain, too, unless the surface effect is so important.
%Values of coefficient $c_1$, $c_2$ depend on the flexibility mechanism through the form of the orientational entropy.
%For example, for a freely-jointed chain $c_1=3.25$, $c_2=4.86$, while for a worm-like-chain the transition occurs at a slightly higher concentration and the coexistence region is narrower $c_1=10.48$, $c_2=11.39$.
%It is interesting to note that the coexistence under consideration indicates the possibility of an ``intra-chain" separation.
%In the present discussion on the scaling level, we do not take seriously such difference.
The free energy in the nematic state is dictated by the loss of the orientational entropy, thus, evaluated to be proportional to the number of the statistical segment
\begin{eqnarray}
\frac{F}{k_BT} \simeq \frac{F_{ori}}{k_BT} \simeq p^{-1} N
\end{eqnarray}
%The coefficient $c_0$ increases with the concentration (degree of the ordering).
This is just a crude approximation since the degree of the ordering increases (even rather slowly) with the concentration.
At much higher concentration, this would not be certainly correct due to higher order terms.

(IV) {\it Ideal chain regime}: $\max\{p, p^{-1}N\}<D/b<(pN)^{1/2}$

We now consider the meaning of the second condition $p^{-1}L<D$ necessary for the bulk semidilute solution analogy.
To do so, let us compare the correlation length of the concentration fluctuation with the cavity size.
In the fluctuating semidilute regime, the condition $D>\xi$ (eq. \ref{xi_fluct_stiff}) is equivalent to trivial condition for the confinement $D<R$.
On the other hand, in the mean-field semidilute regime, $D>\xi_{MF}$ (eq. \ref{xi_mean_field}) leads to nontrivial condition $D<p^{-1}L$.
Note that for chains not too long ($N<p^3$), this length $p^{-1}L$ is still smaller than the natural chain size in bulk $R=(lL)^{1/2}$.
Here the correlation length calculated in bulk exceeds the system size, indicating that the confined chain is no longer analogous to the bulk semidilute solution at $D>p^{-1}L$.
This regime can be regarded as a critical point analogue in a finite confined system (``critical region") and never expected for a flexible chain.
Approaching to the line $D=p^{-1}L$ from the mean-field semidilute regime, the bulk region decreases and finally the whole system becomes nonuniform.
The free energy associated with the nonuniformity is no longer regarded as the surface term and the confinement free energy is evaluated from the integral of eq.~(\ref{DeltaF_SCF_formal}) by replacing $\zeta$ with $D$.
\begin{eqnarray}
\frac{F}{k_BT} = \frac{\Delta F}{k_BT} = A_1 \left( \frac{b}{D} \right)^2 pN + A_2 \left( \frac{b}{D} \right)^3 N^2
\end{eqnarray}
($A_1$ and $A_2$ are numerical coefficients of order unity).
In the present case with $D>p^{-1}L$, the first term (arising from the reduction in the conformational entropy) is dominant, i.e., excluded volume interactions are disregarded and the confinement free energy is approximated by that of an ideal chain trapped in a cavitiy $F/(k_BT) \simeq (b/D)^2pN$.
%Approaching to the line $D=p^{-1}L$ from the mean-field semidilute regime, the volume interaction decreases and becomes the same order as the reduction in the conformational entropy due to the confinement.
%\begin{eqnarray}
%\frac{F_{ent}}{k_BT} \simeq \frac{N}{g_D} \simeq \left( \frac{b}{D} \right)^2 pN
%\end{eqnarray}
%where the number of free monomers $g_D$ between successive collisions with the cavity wall is obtained from $ag_D^{1/2}=D$.
%This is, in fact, the confinement free energy of an {\it ideal chain}.
%In the present ideal chain regime, excluded volume interactions are totally disregarded and the confinement free energy is dominated by this entropic contribution $\Delta F \simeq F_{ent}$.

(V) {\it Bending regime}: $D/b<\min\{p,N\}$

The behaviours of the confined chain should be strongly dependent on the flexibility mechanism once the cavity size becomes smaller than the Kuhn segment length.
In particular, it is obvious that a freely-jointed chain cannot be placed inside such a small space.
We consider here a worm-like-chain, which is suitable as a model of DNA packing inside virus capsid.
A worm-like-chain possesses a uniform elastic modulus along the chain contour $\kappa = k_BT l/2$.
When such a chain is confined in a small cavity with size $D<l$, each chain section is forced to retain the curvature on the order of $D^{-1}$.
In average, the bending energy is almost uniformly distributed along the chain.
Thus, our estimate for the confinement free energy is
\begin{eqnarray}
\frac{F}{k_BT} \simeq \frac{F_{bend}}{k_BT} \simeq \frac{\kappa}{k_BT} \frac{L}{D^2} \simeq \left(\frac{b}{D}\right)^2 pN
\label{F_bend}
\end{eqnarray}
The loading force is not length dependent, which is supported by the result of computer simulation if not the segment concentration is too high ($\phi \lesssim 0.2$)~\cite{Kindt}.
Again, at high concentrations, segmental interactions become important so that the confined chain may exhibit some structural order and eq. (\ref{F_bend}) is no longer applicable.
There are several studies for such a situation in connection to the problem of DNA packing inside virus capsid~\cite{Kindt, Odijk1, Odijk2}.

%One cannot gain any information about the structural order from above discussion, but it would be legitimate to assume that bent segments exhibits local orientational ordering in high enough concentrations.
%But it should be noted again that at high concentrations, segmental interactions may become important so that eq. (\ref{F_bend}) needs some correction.
%A bulk theory for the isotropic-nematic transition can not be directly applied to that for the present situation.
%Thus, the orientational distribution of strongly bent segments is not clear.
%But it would be legitimate to assume that segments are oriented in the dark area (high density)  and not ordered in the white area (low density) in Fig. 

\section{Discussion}
\label{Discussion}
In this section, we shall discuss some generalization of the obtained results.
We have assumed the confinement in a spherical container only.
This is just for the simplicity and the generalization to asymmetric geometry is straightforward.
For instance, let us compare a chain confined in a rectangular box of volume $V=L_1 L_2 L_3$ ($L_1 < L_2 <L_3$) and that in a sphere with equal volume (radius $D = (3V/4\pi)^{1/3}$).
Since the link concentrations are the same, the confinement free energies are also the same in the leading order in regime (I) and (II), and the slight difference just comes from the surface term $\Delta F$.
We expect that this explains the subtle shape effect of the capsid found in a recent simulation that a sphere packs more quickly and ejects more slowly a flexible chain than an ellipsoid~\cite{Yoemans}.
On the crossover from the confined to the unconfined situation, there appears a WCR between SCR ($\xi < L_1$) and bulk ($R < L_1$), where the correlation length is set by the smallest size ($\xi = L_1$), i.e., a chain confined in a slit.
The border between regime (II) and (IV) is given by $L_2L_3/L_1  \simeq L/p$, and in regime (IV) the entropic contribution as a dominant term is set by the smallest size, i.e., $F/(k_BT) \simeq (b/L_1)^2pN$.

Effect of the solvent quality is also easily included by adding the reduced temperature term in the second virial coefficient, i.e., $B=v \tau$, provided that the corresponding semidilute solution stays in the one-phase region.
However, for a polymer with large $p$, the (intra-chain) phase separation and/or the transition to highly ordered nematic states would occur above $\theta$ point as expected from the theory of the semiflexible polymer solution~\cite{RedBook,Gresberb_Khokhlov_LC}.

\section{Summary}
\label{Conclusion}
We have discussed static properties of a long linear polymer confined in the closed cavity.
Guided by an analogy between a long confined chain in SCR and semidilute solutions, we have found five distinct regimes for a semiflexible chain ($p \gg 1$) depending on the contour length and the cavity size.
Three of them have semidilute solution analogues, but other two are intrinsic to a confined chain.
We have not aimed to describe high concentration states (liquid crystalline and bending regimes), since properties there are not universal and more elaborate models are required~\cite{Kindt, Odijk1, Odijk2}.
For other regimes, the obtained results are rather general thanks to the universality of non-concentrated polymer solutions.

Approaching to the flexible chain limit ($p \rightarrow 1$), all the regimes merge into a fluctuating semidilute regime, in which the confinement free energy is given by $F \sim N^{9/4}$.
Flexible chains such as polyethylene glycol belong to this.
Note that for a confined chain, the presence of a slight segmental attraction leads to a semidilute in $\theta$-solvent regime ($F \sim N^{3}$).

Now we discuss the experimental accessibility of the present predictions.
As an example of semiflexible chains, let us take up DNA molecules.
The Kuhn length and the width of the double strand are $l \simeq 100$ nm and $b \simeq 2$ nm, respectively, in usual aqueous conditions, thus, the aspect ratio $p \simeq 50$ (note that the electrostatic repulsion results in some thickening at low ionic strength).
First, we see that a fluctuating semidilute regime requires very long DNA ($L \gtrsim b p^3 \simeq 250 \mu$m).
Although many natural DNA in biological origin is long enough to satisfy this, it is difficult to treat such long chains in a reliable manner.
Therefore, other regimes are more easily accessible and relevant to most experiments.
For moderate (but still rather long) chains ($5 \mu$m $\lesssim L \lesssim 250 \mu$m), a variety of regimes are expected depending on the cavity size.
For example, if T4 DNA ($L\simeq 56 \mu$m) is trapped in the cavity with size ca. $1 \mu$m, a mean-field semidilute regime ($F \sim N^{2}$) is most probable, while an orientationally ordered state may be observed in the small cavity of a few submicron size. 
In the case of $\lambda $ DNA ($L\simeq 16 \mu$m), an ideal chain regime ($F \sim N$) is expected as long as the cavity size exceeds a few submicron, etc.
For shorter chains ($L \lesssim 5 \mu$m), we expect an ideal chain regime, or a bending regime depending on the relative size ratio between $l$ and $D$.

Although qualitative (all the numerical coefficients of order unity are not determined), we believe that the present discussion provides a useful guide for emerging techniques operating long polymers, such as DNA, in $nm \sim \mu m$ scale spaces, where molecules may be highly compressed geometrically~\cite{DNA_partitioning_model_pore} or by external field~\cite{Ichikawa}.
It is also relevant to phase behaviours of composite soft matter systems such as a mucroemulsion + polymer system~\cite{EPL_Nakaya}.
In certain situations (e.g., a polymer compressed by electric filed against a wall), a nonuniform link distribution should be taken into account~\cite{CompressionDNA,SakaueElectrophoresis}.

\begin{acknowledgments}
Comment and advise from T. Ohta was very helpful to brush up the manuscript. This research was supported by JSPS Research Fellowships for Young Scientists (No. 01263).
\end{acknowledgments}

\appendix
\section{Calculation of $\Delta F$}
\label{DeltaF}
Here we review briefly what is needed for our discussion.
Consider a long semiflexible chain confined in a spherical cavity (radius $D$).
The solvent is assumed to be athermal and the cavity wall is repulsive (no adsorption).
The local link concentration is almost uniform ($c(r)= c \simeq N/D^3$) within the cavity, but drops the near the wall toward $c(r=D)=0$.
If the local fluctuation is negligible, the concentration profile can be obtained by self-consistent field calculation~\cite{deGennes}.
The self-consistent equation is
\begin{eqnarray}
\left( -\frac{a^2}{6}\nabla^2  + \frac{U_{SCF}(r)}{k_BT} \right) \psi(r) = \epsilon \psi(r)
\end{eqnarray}
where $\psi(r)$ is the grand state eigen function of the propagator (normalized as $c(r) = |\psi(r)|^2$) and the potential is
\begin{eqnarray}
\frac{U_{SCF}({\mathbf r})}{k_BT} = v |\psi(r)|^2
\end{eqnarray}
By imposing the boundary condition $\psi (D/2) =0$ and $\psi(r) =c^{1/2}$ far away from the wall, one finds the grand state eigen value $\epsilon = vc$ (consistent with the free energy eq.~(\ref{F_meanfield})) and the profile
\begin{eqnarray}
c(r) =  c  \ \tanh^2\left( \frac{D-r}{\zeta}\right) \qquad (r\le D)
\label{c_profile}
\end{eqnarray}
where $\zeta = b (p/3 \phi)^{1/2}$, which is, in essence, the correlation length ($\zeta \simeq \xi_{MF}$ from eq.~(\ref{xi_mean_field})), signifies the range for the depletion layer near the wall.
The essentially same form as eq.~\ref{c_profile} is given in ref.~\cite{deGennes} for the monomer concentration profile of semidilute solution near the repulsive wall.
The surface free energy is evaluated as
\begin{eqnarray}
\frac{\Delta F}{k_BT} = \int dS \int_{D-\zeta}^{D} dr \left\{\frac{a^2}{6}(\nabla \psi(r)) ^2 + \frac{1}{2}v|\psi(r)|^4 \right\} \label{DeltaF_SCF_formal} \\
\simeq \frac{bp^2 D^2}{\xi_{MF}^3}
\label{DeltaF_SCF}
\end{eqnarray}
The same result is obtained from a slightly different way.
The surface free energy is equal to the work required to create a depletion layer against the osmotic pressure.
In a mean-field approximation, the osmotic pressure is given by
\begin{eqnarray}
\Pi = k_BT \frac{\phi^2}{b^3} + \cdots
\end{eqnarray}
Therefore, $\Delta F \simeq \int dS \ (\Pi \times \xi_{MF})$, which is shown to be equivalent to eq.~(\ref{DeltaF_SCF}). 

The latter approach can be conveniently applied to other situations such as a $\theta$ solvent and a fluctuation dominant regime.
In the semidilute in $\theta$-solvent regime, the osmotic pressure is approximated as $\Pi \simeq k_BT \phi^3/b^3$ (dominated by third virial term).
The surface free energy for a flexible chain ($p=1$) is, thus, $\Delta F /(k_BT) \simeq (D\phi/b)^2 \simeq (D/\xi_{MF})^2$ (eq.~(\ref{DeltaF_theta})).
In the fluctuating semidilute regime, $\Pi \simeq k_BT/\xi^3$, where one should notice the correlation length is obtained from the scaling argument (eq.~(\ref{xi_good}),(\ref{xi_fluct_stiff})).
This leads to $\Delta F /(k_BT) \simeq (D/\xi)^2=\phi^{3/2}(p \tau)^{1/2}(D/b)^{2}$ (eq.~(\ref{DeltaF_good})).

\end{document}